\documentclass{article}
\usepackage{hiph-preprint}
\usepackage{graphics}
\usepackage{amssymb}
\usepackage{epsfig}

\volnumber{22} \issuenumber{1} \edyear{2005}                             
\frompage{000} \topage{000}                                              
\recrevdate{18 November, 2005}                                              
\def\rightmark{Nuclear effects on open charm production in p+A reactions}
\def\leftmark{I.~Vitev, T.~Goldman, M.~B.~Johnson and J.~W. Qiu}
 
\title{Nuclear effects on open charm production \\ in p+A reactions} 
\authors{ 
{Ivan Vitev$^1$, T. Goldman$^1$, Mikkel Johnson$^1$ 
and Jian-Wei Qiu$^{2}$ %
\index{Vitev, I.} 
\index{Goldman, T.} 
\index{Johnson, M.~B.} 
\index{Qiu, J.~W.} 
}\\[2.812mm]
{\normalsize
\hspace*{-8pt}$^1$ Los Alamos National Laboratory, 
Theoretical Division and Physics Division,\\ 
Los Alamos, NM 87545, USA\\[0.2ex] 
\hspace*{-8pt}$^2$ Iowa State University,
Department of Physics and Astronomy,\\ 
Ames, IA 50011, USA
}}
 
\abstract{
We calculate and 
resum the nuclear enhanced power corrections from the final 
state partonic scattering in nuclear matter to open charm 
production and 
correlations. 
In p+A reactions, we find that single and double inclusive D mesons 
can be suppressed as much as the neutral 
pions 
from the dynamical high twist shadowing. Effects of 
initial state energy loss in p+A collisions are also investigated 
and may lead to significantly weaker transverse momentum dependence 
of the nuclear suppression. 
}


 
\makeindex
\begin{document}
 
\maketitle

\section{Introduction}\label{intro}
A good probe of the dense nuclear medium created in relativistic
heavy ion collisions should be sensitive to 
its dynamical scales~\cite{Djordjevic:2004nq,hq,Qiu:2003vd},  
while it can be cleanly
measured experimentally and reliably calculated theoretically.
Because of  color confinement, only hard probes
can be reliably calculated in the perturbation theory of Quantum 
Chromodynamics.
On the other hand, typical dynamical scales of the dense medium 
produced in p+A and A+A reactions are of the order of 
hundreds of MeV,  which is much smaller than the scale of a hard 
probe and therefore non-perturbative.
Therefore, an ideal probe should be not only ``hard'' but also 
sensitive to this ``soft'' physics.  Open charm production has 
the potential to be an ideal 
probe because of two distinctive scales of the open charm meson:
the charm quark mass ($m_c \sim 1.5$~GeV, a relatively 
hard scale) and the binding energy ($\sim M_D-m_c\sim$ hundreds 
MeV)~\cite{hq}. This contribution summarizes part of our
study of nuclear effects on D meson production and 
D-meson-triggered correlations in p+A reactions at RHIC.

\section{Coherent power correction and initial state 
energy loss}\label{present}

We calculate the open charm ($D^0$ and $D^+$ meson) production cross 
sections to lowest order in perturbative QCD, using standard
CTEQ6 parton distribution 
functions and fragmentation functions derived by Braaten et al. in
heavy quark effective field theory~\cite{Braaten:1994bz}. The contribution
from  scattering of the radiatively generated charm is 
included in our calculation~\cite{hq,Qiu:2004da}.  We find 
that for inclusive single charm production $c+g \rightarrow c+g$ and 
$c+q(\bar{q}) \rightarrow c+q(\bar{q})$ dominate over gluon fusion 
$g+g \rightarrow c+\bar{c}$. 
The $g+g \rightarrow c+\bar{c}$  process 
controls charm-anticharm pair production. 
More specifically, the 
cross sections are calculated as follows~\cite{hq}:
\begin{eqnarray}
\label{single}
&& \hspace*{-1.3cm}  \frac{ d\sigma^{D^0+D^+}_{NN} }{ dy_1  d^2p_{T_1}  }   =   
K \sum_{abcd} \int dy_2 \int 
\frac{d z_1 }{z_1^2} D_{(D^0+D^+)/c}(z_1) 
   \frac{\phi_{a/N}({x}_a) \phi_{b/N}({x}_b) }{{x}_a{x}_b} \,
\frac{\alpha_s^2}{{S}^2 }  |\overline {M}_{ab\rightarrow cd}|^2   \;, \; \\
&& \hspace*{-1.3cm} \frac{ d\sigma^{(D^0+D^+)-(\bar{D}^0+D^-)}_{NN} }
{ dy_1  dy_2 d p_{T_1}  d p_{T_2} } 
= K \! \sum_{abcd} \!  
2\pi  \int_{\cal D} 
\frac{d z_1}{z_1}  D_{(D^0+D^+)/c}(z_1)  D_{(\bar{D}^0+D^-)/d}(z_2)
 \nonumber \\
  && \hspace*{2.5cm}  \times \frac{\phi_{a/N}({x}_a) \phi_{b/N}({x}_b) }{{x}_a{x}_b} \,
\frac{\alpha_s^2}{{S}^2 }  
|\overline {M}_{ab\rightarrow cd}|^2   \;. 
\label{double}
\end{eqnarray}

The nucleus-induced  modification to the cross sections measured in $A+B$ 
collisions relative to the binary scaled 
$p+p$ result is identified through the ratio   %
\begin{equation}
 R^{(n)}_{AB}  =  \frac{d\sigma^{h_1 \cdots h_n}_{AB} / 
dy_1 \cdots dy_n d p_{T_1} \cdots  d p_{T_n}} 
{\langle N^{\rm coll}_{AB} \rangle\, d\sigma^{h_1 \cdots h_n}_{NN} / 
dy_1 \cdots dy_n d p_{T_1} \cdots  d p_{T_n}} \; .
\label{multi}
\end{equation}
$\langle N^{\rm coll}_{AB} \rangle$ in Eq.~(\ref{multi}) is often 
evaluated from an optical Glauber model. Here $h_i$ can be any 
combination of final state hadrons, for example $h_1 = D^0+D^+$, 
$h_2 = \bar{D}^0+D^- $.

The effects of coherence can be cleanly studied in deep inelastic 
scattering~\cite{Qiu:2003vd}. If the longitudinal momentum transfer 
becomes small, the probe will interact simultaneously  
with more than one nucleon. In a frame where the longitudinal size of 
the exchange virtual photon is given by $l_c = 1/xP$ the Lorentz
contracted nucleon has longitudinal size $2r_0/\gamma = 2r_0/(P/m_N)$.   
The critical value for the onset of coherence then 
reads $ x_N = 1 / (2 r_0 m_N) \sim 0.1$~\cite{Qiu:2003vd}.
In our work~\cite{hq}, we combined the computation of the
dynamical mass, generated by multiple final state scattering, 
for  heavy and light quarks~\cite{Qiu:2003vd}
with the observable effects of 
high twist modification of hadron production in
p+A reactions~\cite{Qiu:2004da}.   
Specifically, isolating the  small $x_b$ dependence 
in Eqs.~(\ref{single}) and~(\ref{double}) into $F_{ab\rightarrow cd}(x_b) 
 =  \phi_{b/N}(x_b) |M_{ab\rightarrow cd}|^2/x_b$, we resum 
all of the high twist nuclear-enhanced 
corrections in the regime $|\hat{t}| \ll |\hat{s}|,\, |\hat{u}|$.
We find
\begin{eqnarray}
 F_{ab\rightarrow cd}(x_b) &\Rightarrow & 
F_{ab\rightarrow cd}\left( x_b\left[ 1+ C_d \frac{\xi^2 (A^{1/3}-1)}
{-\hat{t} + m_d^2}  \right] \right)\;.
\label{finalshift}
\end{eqnarray}
In Eq.~(\ref{finalshift}) $C_d = 1\; (9/4)$ for quarks (gluons), 
respectively. When $ m_d \rightarrow 0 $  we recover the result 
of~\cite{Qiu:2004da}. (Note that $m_d$ can be the mass of the charm
quark.)

\begin{figure}[t!]
\vspace*{0cm}
\includegraphics[width=2.3in,height=3.in]{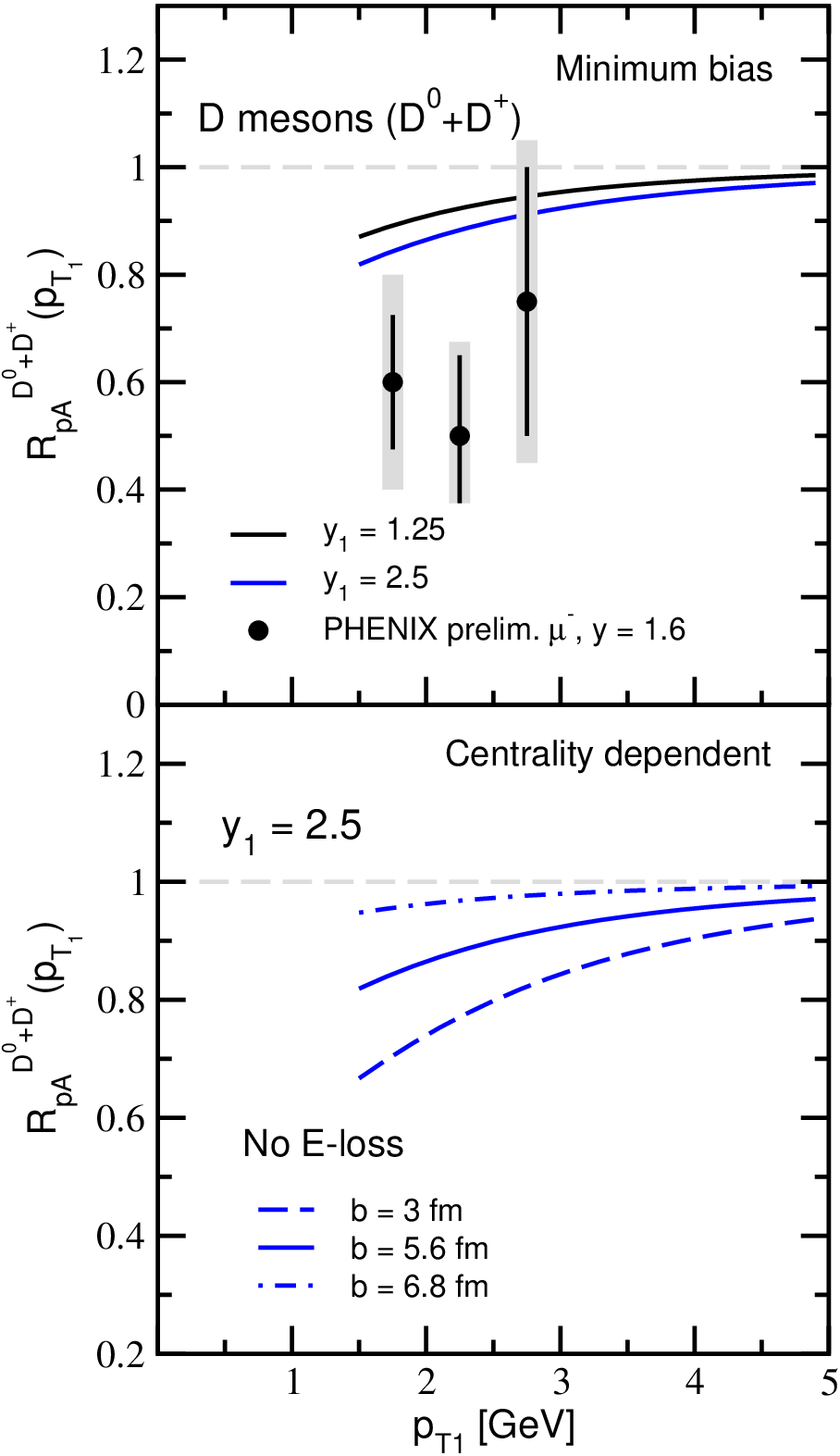}
\hspace*{0.2in}
\includegraphics[width=2.2in,height=3.in]{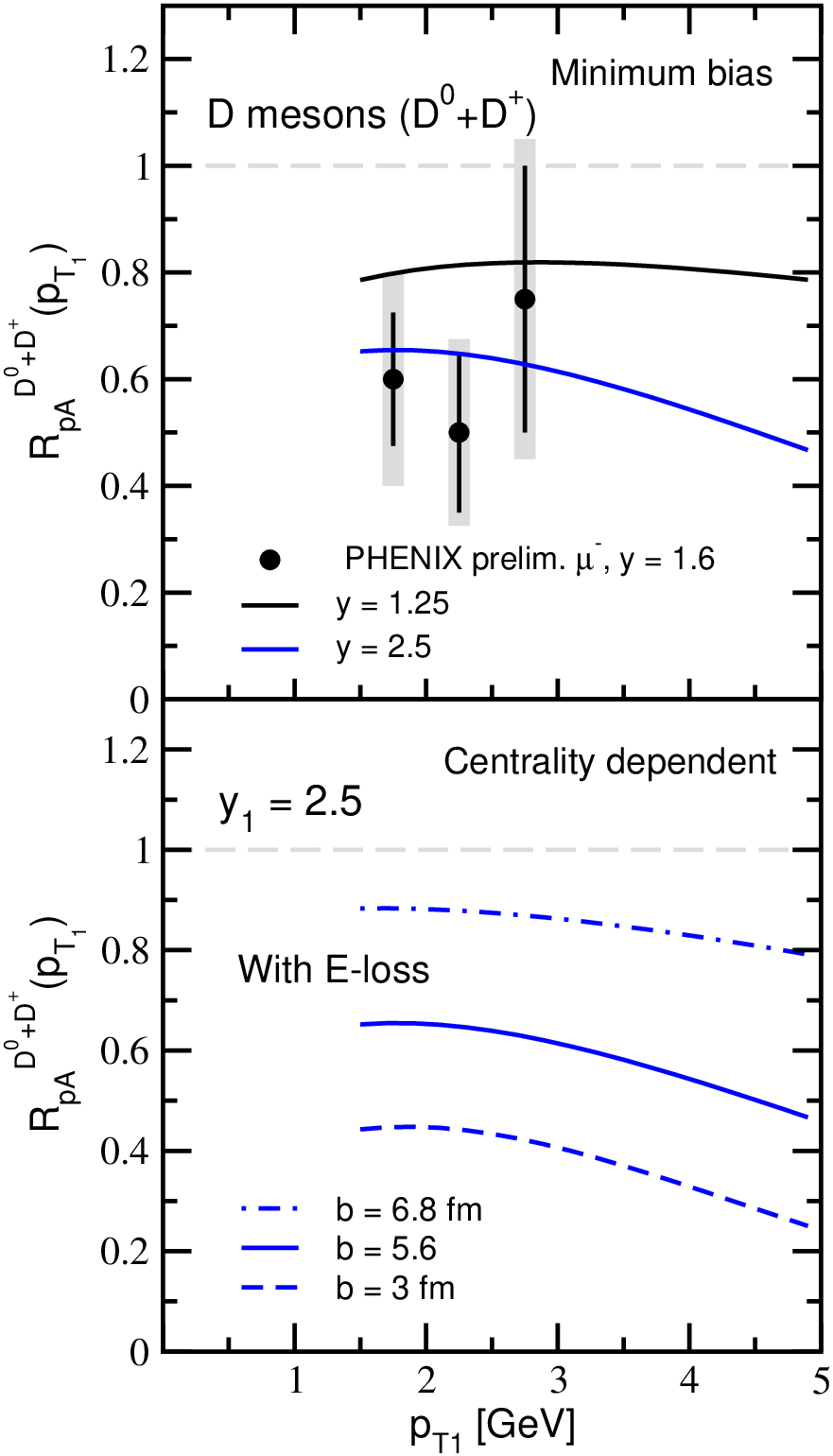}
\vspace*{0cm}
\caption[]{Left panels: coherent power correction effects on 
inclusive D meson production at two different rapidities, 
$y = 1.25$ and $y = 2.5$, in d+Au collisions at 
$\sqrt{s_{NN}}=200$~GeV.  Right panels: the effect of 
power corrections and energy 
loss on D meson production. Bottom panels show the centrality
dependence of the nuclear effects for  $y = 2.5$ only.}
\label{fig1}
\end{figure}

The left panel of Figure 1  shows the nuclear modification
$R^{D^0+D^+}_{pA}(p_{T_1})$ for the single inclusive $D^0+D^+ $ production 
for two different rapidities $y_1 = 1.25$ and  $y_1 = 2.5$. The 
nuclear effect is similar to that for light pions~\cite{Qiu:2004da}.
The centrality dependence of high twist shadowing is also shown. 
Up to rapidity $y_1 = 2.5$, the effect of power corrections is 
relatively small. Preliminary PHENIX data~\cite{XRWang} 
on the nuclear modification for muons, presumably 
coming from heavy quark 
decays, is shown for comparison. The nuclear effect appears to be 
larger than the one predicted by power corrections. Similar 
results for $R^{(D^0+D^+)-(\bar{D}^0+D^-)}_{pA}(p_{T_1},p_{T_2})$ in 
$(D^0+D^+) - (\bar{D}^0+D^-)$ correlations 
as a function of the rapidity gap, $y_2 - y_1$, and centrality  are
given in the left panels of Figure 2.

Additional nuclear attenuation may come from the energy loss
of incoming partons in cold nuclear matter. Initial state medium 
induced radiation, when followed by a hard scattering in the 
nuclear environment, must be clarified in the pQCD formalism 
prior to precise quantitative applications in cold nuclear matter.
Thus, we are motivated
to carry out only a phenomenological study of $\Delta E$ effects
by the known incoherent Bertsch-Gunion type radiation, which scales 
with the path length through the nucleus, $L$~\cite{Gunion:1981qs} 
\begin{equation}
\frac{\omega dN^{(1)}_g}{d \omega d^2{\bf k}_\perp} \sim 
 \frac{\alpha_s}{\pi^2}
\frac{{\bf q}_1^2} {{\bf k}_\perp^2({\bf k} - {\bf q}_1)_\perp^2} \; ,
\qquad \Delta E ^{(\bar{n})} \sim C_F \frac{\alpha_s}{\pi}
E \frac{L}{\lambda_g} \ln \frac{\mu^2}{\Lambda_{QCD}^2}  \; ,
\label{BGrad}
\end{equation}
where $\bar{n} = L/\lambda$ is the mean number of scatterings.
If the incoming partons lose a fraction of their energy,
the effective rapidity of the collision is shifted in the 
direction of the propagation of the nucleus (backwards for 
RHIC kinematics). For minimum bias collisions, we use 
a rapidity shift of $ \Delta y = 0.25$. This is not incompatible 
with the nuclear modification of light hadrons~\cite{eloss} 
and the rapidity asymmetry observed in low energy p+A 
reactions ($\sqrt{s}=20$~GeV), where effects of coherence 
are negligible~\cite{hq}. The variation of  $ \Delta y $
with centrality is given by the nuclear thickness $T_B(b)$. 
 The right panels of 
Figures~1 and~2 show the consequence 
of such an effective implementation of the energy loss. Firstly,
the nuclear suppression  has a weak $p_T$ dependence 
in both single inclusive open charm production 
and charm anticharm correlations. Secondly, the effect
on  $R^{D^0+D^+}_{pA}(p_{T_1})$, 
$R^{(D^0+D^+)-(\bar{D}^0+D^-)}_{pA}(p_{T_1},p_{T_2})$ 
can be large, compared to power corrections for
transverse momenta larger  than few GeV. In particular, the 
right panel of Figure 1 shows suppression much more compatible 
with the preliminary PHENIX measurements. A constant 
rapidity shift may fail if we go to either high 
$p_T$ or forward rapidity since we do expect 
energy dependence of the fractional energy loss~\cite{eloss}. 
This points to the 
necessity of careful studies of energy loss effects in cold 
nuclear matter.

\begin{figure}[t!]
\vspace*{0cm}
\includegraphics[width=2.3in,height=3.in]{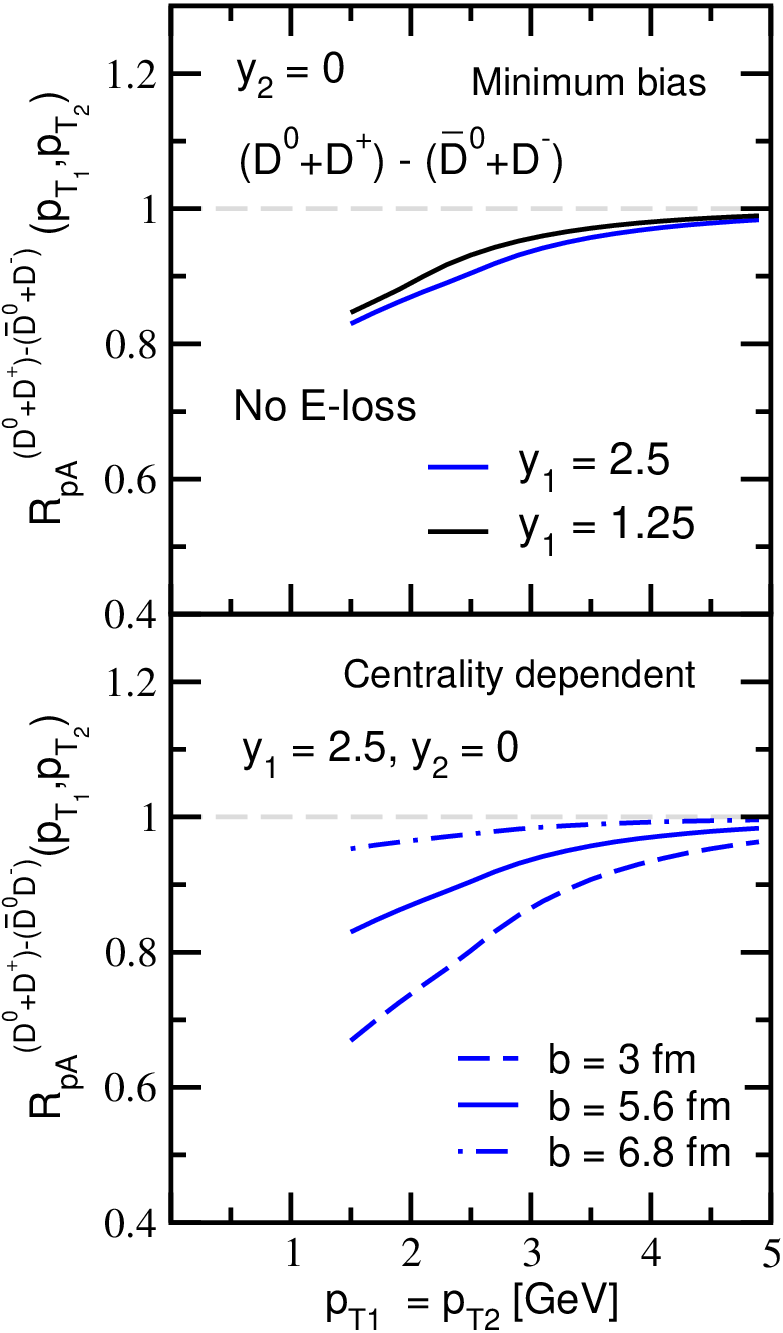}
\hspace*{0.2in}
\includegraphics[width=2.2in,height=3.in]{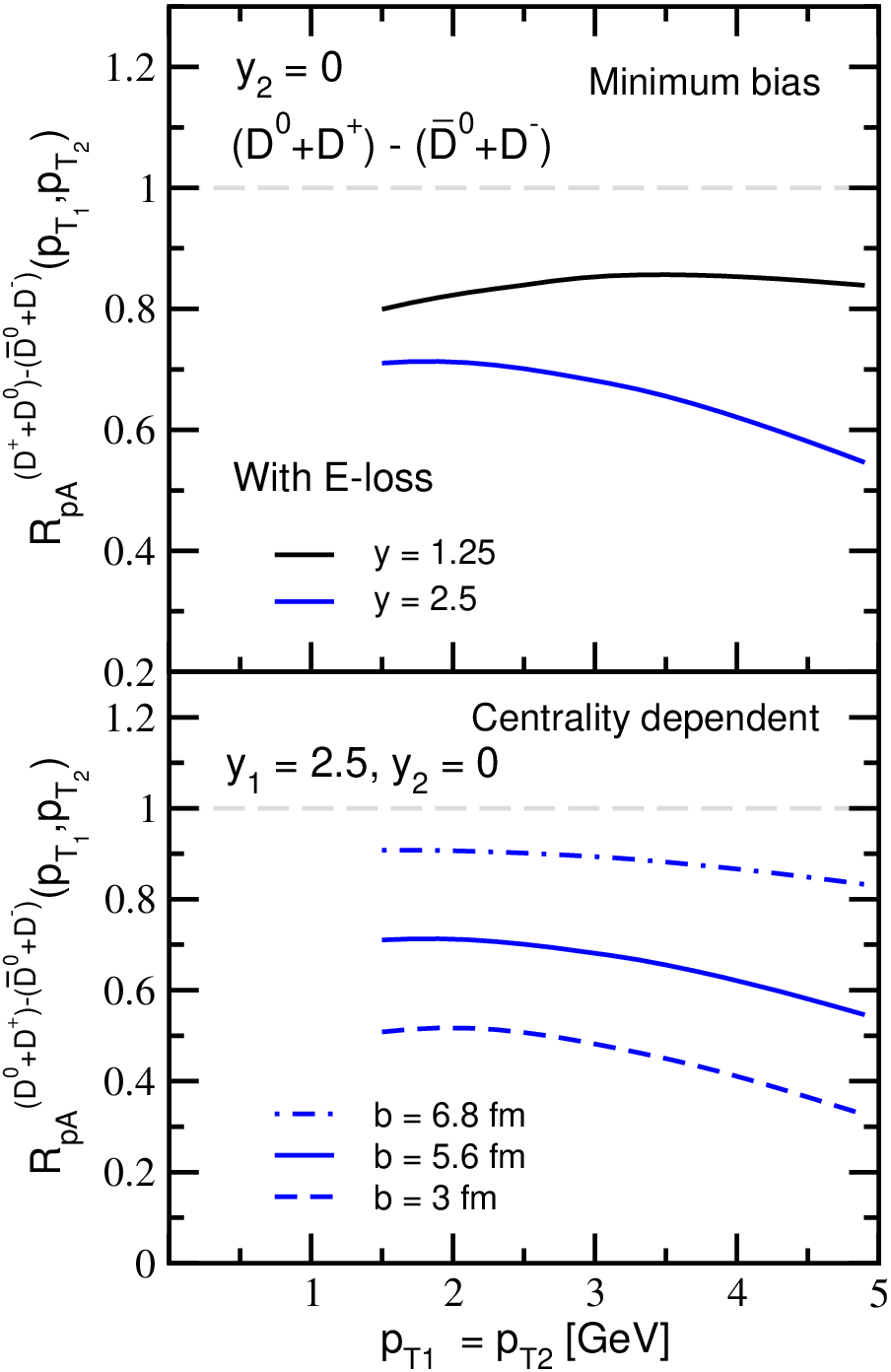}
\vspace*{0cm}
\caption[]{Nuclear attenuation 
from power corrections (left panels) 
and power corrections and energy loss (right panels)
in correlated back-to-back D and anti D meson production in 
d+Au at RHIC. The trigger D meson rapidities are 
$y_1=1.25$ and $y_1=2.5$  and the associated 
anticharm mesons are measured at $y_2=0$. Centrality 
dependence is shown for $y_1=2.5$ only.}
\label{fig2}
\end{figure}

We note, finally, that it is a characteristic of both power 
corrections and initial state energy loss that they reduce
the rate of the hard scattering processes, thus 
similarly affecting single and double inclusive hadron 
production.

\section{Conclusions}\label{concl}
In summary, we have presented the first calculation of coherent 
resummed QCD power corrections to open charm production and 
open charm triggered back-to back correlations~\cite{hq}. We 
find that the nuclear suppression is qualitatively similar 
to that for light hadrons~\cite{Qiu:2004da}. Initial state
energy loss effects have also been phenomenologically 
incorporated and shown to yield a different transverse momentum 
behavior of the nuclear suppression. We expect our theoretical 
results will guide the experimental determination of the relevant
many body QCD effects that control the dynamics of forward
rapidity hadron production in p+A 
reactions~\cite{XRWang,Adler:2004eh}. They will also provide 
the much needed baseline for precision heavy quark QGP 
tomography~\cite{Djordjevic:2004nq} at forward rapidities.

\section*{Acknowledgments}
This research is supported in part by the US Department of Energy  
under Contract No. W-7405-ENG-36,  Grant No. DE-FG02-87ER40371 and by 
the J. Robert Oppenheimer Fellowship of the Los Alamos National 
Laboratory. 

\vfill\eject
\end{document}